\documentstyle[12pt]{article}
\begin{document}
\thispagestyle{empty}
\begin{center}
\LARGE \tt \bf {Structure formation in anisotropic Einstein-Cartan cosmologies}
\end{center}

\vspace{5cm}

\begin{center}
{\large By L.C. Garcia de Andrade\footnote{Departamento de F\'{\i}sica Te\'{o}rica - Instituto de F\'{\i}sica - UERJ 

Rua S\~{a}o Fco. Xavier 524, Rio de Janeiro, RJ

Maracan\~{a}, CEP:20550-003.}}
\end{center}

\begin{abstract}
Density linear perturbations in Einstein-Cartan two fluid cosmologies where the outer model is an isotropic Friedmann solution with closed model while the inner model is a flat anisotropic Einstein-Cartan (EC) cosmology with shear are computed by taking into consideration superhorizon situations.This situation is in agreement with the present isotropy of the Cosmic Background Radiation where the anisotropic EC model evolves into an isotropic general relativistic model.Galaxy formation is discussed based on this model.Another type of two fluid is considered where the inner solution is a pressureless Kasner type solution while the outer solution is a de Sitter solution with spin and torsion.
\end{abstract}

\newpage
\section{Introduction}

Early Perko,R.Matzner and L.C.Shepley (PMS) \cite{1} following the work by Lifichitz and Khalatnikov \cite{2} on perturbations in isotropic cosmological models have investigated the formation of galaxies and perturbations of anisotropic general relativistic cosmological models. They showed that the density perturbations would demand that the formation of galaxies possess a time dependence of $t^{0}$ to $t^{\frac{8}{3}}$ depending on the model used.In this report I show that by making use of a two fluid system one general relativistic isotropic closed model and the other component of the fluid woul be a  anisotropic flat cosmological EC model with shear and spin polarization only along one of the axis ,which was proposed by Kuchowicz years ago \cite{3,4}.We show that our model fits very well within the PMS model when the spin-torsion vanishes since it gives a density perturbation proportional to $t^{\frac{4}{3}}$ which is bigger than the isotropic value of $t^{\frac{2}{3}}$ which gives a not so higher growth of inhomogeneities as the one obtained in PMS and is even higher than the approximated short-wave limit of $t^{\frac{5}{3}}$ proposed in PMS.Meanwhile the spin-torsion contribution to density perturbations is $t^{-\frac{4}{3}}$ in the radiation phase and $t^{-2}$ in the matter dominated phase.This work generalizes recent work we make on density perturbations in the isotropic case \cite{5}.The contribution of spin-torsion fluctuations is only important before the decoupling between matter and radiation and is redshifted at present age of the Universe.The need of considering anisotropic models in EC gravity where a preferred spin polarization direction is considered was first pointed out by Kuchowicz \cite{6}.The contribution of spin-torsion is therefore important in the Early Universe \cite{7}.The strongest argument to use anisotropic models is that torsion contributions to galaxy formation occurs at the early stages of the Universe when the cosmological model must be anisotropic \cite{1}.Finally another two fluid example is given where the inner solution is a Kasner solution without spin and torsion while the outer solution is a de Sitter
solution with spin and torsion.The density perturbation associated with this model yields terms such as $t^{2}<t^{\frac{8}{3}}$ and the spin-torsion introduces a contribution such as $t^{3}>t^{\frac{8}{3}}$ which yields a density perturbation higher than the usual perturbations found in General Relativistic Anisotropic Cosmology. 
\section{The Model}
Let us now consider the flat anisotropic EC model given by the metric 
\begin{equation}
ds^{2}=dt^{2}-X^{2}(t)dx^{2}-Y^{2}(t)(dy^{2}+dz^{2})
\label{1}
\end{equation}
where the density of angular momentum is given by $S_{ij}^{k}=u^{k}S_{ij}$ where 
$u^{i}$,${i,j=0,1,2,3}$ is the four-velocity of the fluid.The axis-x is the spin alignement of polarization and $S_{23}=-S_{32}$.The expansion is given $\frac{\dot{a}}{a}=\frac{1}{3}{\theta}$ and the relation of the cosmic scale $a$ with the metric components X and Y is $a(t)=(X(t)Y^{2})^{\frac{1}{3}}$.With this change one can tranform the field equations in X and Y \cite{5} to
\begin{equation}
8{\pi}G{\rho}=3(\frac{\dot{a}}{a})^{2}-\frac{1}{3}\frac{C^{2}}{a^{6}}+(4{\pi}GS_{23})^{2}
\label{2}
\end{equation}
and
\begin{equation}
8{\pi}G{\rho}=-2{\frac{\ddot{a}}{a}}-(\frac{\dot{a}}{a})^{2}-\frac{1}{3}\frac{C^{2}}{a^{6}}+(4{\pi}GS_{23})^{2}
\label{3}
\end{equation}
where the shear is given by the expression
\begin{equation}
3^{\frac{1}{3}}|{\sigma}|=\frac{C}{a^{3}}
\label{4}
\end{equation}
and the spin density is
\begin{equation}
S_{23}=\frac{S_{0}}{a^{3}}
\label{5}
\end {equation}
and of course we consider in this model that the spin is conserved.Now let us consider the inner isotropic spintorsion and shear free model given by the equation
\begin{equation}
8{\pi}G{\rho}_{0}=(\frac{\dot{a}_{0}}{a_{0}})^{2}+\frac{1}{a^{2}}
\label{6}
\end{equation}
where the last term reflects the fact that we are considering a closed cosmological model.Performing the difference between the equations (\ref{2}) and (\ref{6}) we obtain the density contrast
\begin{equation}
{\delta}=\frac{{\delta}{\rho}}{{\rho}_{0}}=\frac{1}{8{\pi}G}[\frac{2}{{\rho}_{0}}(\frac{\dot{a}}{a})^{2}-\frac{1}{3}\frac{C^{2}}{a^{6}{\rho}_{0}}+\frac{{4{\pi}GS_{0}}^{2}}{a^{6}{\rho}_{0}}]
\label{7}
\end{equation}
In the radiation dominated phase ${{\rho}_{0}}|_{R}{\alpha}a^{-4}$  and the matter dominated phase ${{\rho}_{0}}|_{M}{\alpha}a^{-3}$ .Substitution of these values into equation (\ref{7}) one obtains 
\begin{equation}
{\frac{{\delta}{\rho}}{{\rho}_{0}}}|_{M}=\frac{1}{8{\pi}G}[2{\dot{a}}^{2}{a}+((4{\pi}GS_{0})^{2}-\frac{1}{3}C^{2})a^{-3}]
\label{8}
\end{equation}
and \begin{equation}
(\frac{{\delta}{\rho}}{{\rho}_{0}})|_{R}=\frac{1}{8{\pi}G}[2{\dot{a}}^{2}{a}^{2}+((4{\pi}GS_{0})^{2}-\frac{1}{3}C^{2})a^{-2}]
\label{9}
\end{equation}
Since the outer model is general relativistic we can take the advantage of the known solution $a{\alpha}t^{\frac{2}{3}}$.Substitution of this value into the equations (\ref{8}) and (\ref{9}) we obtain
\begin{equation}
{\delta}|_{R}=\frac{3}{4{\pi}G}t^{\frac{4}{3}}+[((4{\pi}GS_{0})^{2}-\frac{1}{3}C^{2})t^{-\frac{4}{3}}]
\label{10}
\end{equation}
and
\begin{equation}
{\delta}|_{M}=\frac{1}{8{\pi}G}t^{\frac{2}{3}}+[((4{\pi}GS_{0})^{2}-\frac{1}{3}C^{2})t^{-2}]
\label{11}
\end{equation}
Where we have proved the assertions in the beginning of this report.Another not so simple but more general way to obrtain the density perturbations would be to consider the density perturbations from the EC equations in the X-Y form such that
\begin{equation}
8{\pi}G{\rho}=(\frac{\dot{Y}}{Y})^{2}+2\frac{{\dot{X}}{\dot{Y}}}{XY}+(4{\pi}GS_{23})^{2}
\label{12}
\end{equation}
and
\begin{equation}
8{\pi}Gp=-2{\frac{\ddot{Y}}{Y}}-(\frac{\dot{Y}}{Y})^{2}+(4{\pi}GS_{23})^{2}
\label{13}
\end{equation}
and
\begin{equation}
8{\pi}Gp=-{\frac{\ddot{Y}}{Y}}-{\frac{\ddot{X}}{X}}-\frac{{\dot{X}}{\dot{Y}}}{XY }+(4{\pi}GS_{23})^{2}
\label{14}
\end{equation}
Performing the perturbation on the anisotropic solution around the unperturbed one $Y_{0}$ as $Y=Y_{0}+{\delta}Y$ with the same expressions for X we obtain
\begin{equation}
\frac{{\delta}{\rho}}{{\rho}_{0}}=\frac{1}{4{\pi}G{\rho}_{0}}\frac{\dot{Y_{0}}}{Y_{0}}[{\dot{{\delta}X}}+{\dot{{\delta}Y}}]+4{\pi}G\frac{({\delta}S)^{2}}{{\rho}_{0}}
\label{15}
\end{equation}
which yields
\begin{equation}
\frac{{\delta}{\rho}}{{\rho}_{0}}=\frac{1}{4{\pi}G{\rho}_{0}}\frac{\dot{Y_{0}}}{Y_{0}}[\dot{X}-\dot{X_{0}}+\dot{Y}-\dot{Y_{0}}]+4{\pi}G\frac{({\delta}S)^{2}}{{\rho}_{0}}
\label{16}
\end{equation}
Proceeding just as before we obtain
\begin{equation}
{\delta}|_{M}=\frac{2}{3}t^{\frac{1}{3}}[\dot{X}+\dot{Y}]+4{\pi}Gt^{-2}
\label{17}
\end{equation}
for the matter dominated era and
\begin{equation}
{\delta}|_{R}=\frac{1}{4{\pi}G}(\frac{2}{3}[\dot{X}+\dot{Y}]t-\frac{8}{9}t^{\frac{2}{3}}+4{\pi}Gt^{-\frac{4}{3}})
\label{18}
\end{equation}
Note that the second term on the RHS of this last equation is the general relativistic perturbation.Note that these two last expressions tell us that each direction in the anisotropic model contributes differently to the density perturbations.To see this result in an explicit form and have an account of the role of the contribution of spin-torsion to the density fluctuations we must solve the EC field equations to obtain the expressions for X and Y and to substitute into equations (\ref{17}) and (\ref{18}).A simple solution of the field equation can be obtained as follows.First we must equate equations (\ref{13}) to (\ref{14}) and (\ref{12}) to (\ref{13}) to obtain the following equations
\begin{equation}
2\frac{\ddot{Y}}{Y}+2(\frac{\dot{Y}}{Y})^{2}=2\frac{\ddot{X}}{X}+2\frac{{\dot{X}}{\dot{Y}}}{XY}
\label{19}
\end{equation}
and 
\begin{equation}
-2\frac{\ddot{Y}}{Y}-2(\frac{\dot{Y}}{Y})^{2}=2\frac{{\dot{X}}{\dot{Y}}}{XY}
\label{20}
\end{equation}
Now subtracting equation (\ref{20}) from (\ref{19}) one obtains
\begin{equation}
2\frac{\ddot{X}}{\dot{X}}=-\frac{{\dot{Y}}}{Y}
\label{21}
\end{equation}
taking the approximation that the shear is much small than the other quantities involved we obtain after some algebra $X=t^{\frac{1}{2}}$ and that $Y=2t^{-\frac{1}{2}}$ which substitution into equations (\ref{17}) and (\ref{18}) 
produce terms such as $t^{-\frac{1}{6}}$ and $t^{-\frac{7}{6}}$ into the matter dominated phase density perturbation and terms such as $t^{\frac{1}{2}}$ and $t^{-\frac{1}{2}}$ into the radiation dominated era density perturbation.One notes immeadiatly that there are a decaying and a growing mode in the radiation and matter dominated phases.These terms seems not to be able to explain the galaxy formation.Spin-torsion fluctuations in fact introduce terms which decay faster than terms which already appeared in the General Relativistic results,namely in GR we found for the density perturbations a term which decays as $t^{-1}$ while in Einstein-Cartan cosmology we found a term which decays as $t^{-2}$ in the matter dominated era.In the radiation era the situation is not so bad and a term $t^{-\frac{4}{3}}=t^{-1.33}$ is obtained.This contribution is much closer to the GR one.As a final example let us consider the Kasner metric
\begin{equation}
ds^{2}=dt^{2}-t^{p_{1}}dx^{2}-t^{p_{2}}dy^{2}-t^{p_{3}}dz^{2}
\label{22}
\end{equation}
To simplify we consider that $p_{2}=p_{3}$.The components of Riemannian Einstein tensor are
\begin{equation}
G^{0}_{0}=\frac{1}{2t^{2}}(\frac{{p_{1}}^{2}}{2}-p_{1}p_{2}-{p_{2}}^{2})=8{\pi}G{\rho}_{0}
\label{23}
\end{equation}
and since the fluid is pressureless we have that $G^{a}_{b}=0$ where ${i,j=0,1,2,3}$ and combination of these equations yield
\begin{equation}
p_{2}=\frac{(p_{1}-4)}{5}
\label{24}
\end{equation}
and $p_{2}=-p_{1}$.Substitution of this result into expression (\ref{24}) yields 
$p_{1}=\frac{2}{3}$.From the equation $G^{0}_{0}={\rho}_{0}$ and the values for $p_{a}$ we obtain the matter  density of the unperturbed Kasner model as ${\rho}_{0}=\frac{1}{9t^{2}}$, which is in agreement with the matter density ${\rho} {\alpha} a^{-3}$.The inner model solved we now consider the outer cosmological model as a de Sitter solution with spin and torsion given by 
\begin{equation}
{H_{0}}^{2}+\frac{1}{a^{2}}=8{\pi}G({\rho}-2{\pi}G{\sigma}^{2})
\label{25}
\end{equation}
Where $H_{0}$ is the Hubble.Making use of the spin-torsion value and performing the  difference between the formulas (\ref{24}) and (\ref{25}) we obtain the final expression for the density perturbation in the form 
\begin{equation}
{\delta}{\alpha}{H_{0}}^{2}t^{2}+(D+AH_{0})t^{3}
\label{26}
\end{equation}
as we wish to show.Therefore we may concluding that depending on the anisotropic model we use one obtains values of density perturbation high enough to give a growth in the inhomogeneities due to the presence of spin-torsion fluctuations. 
\section*{Acknowledgements}
I would like to thank CNPq. (Brazil) for partial financial support.


\newpage

\begin{thebibliography}{7}

\bibitem{1} T.E.Perko,R.Matzner and L.C.Shepley,Phys.Rev.D,(1972),6,969.
\bibitem{2} E.Lifchitz and I.M.Khalatnikov,Advan.Phys.12,(1963)185.
\bibitem{3} B.Kuchowicz,J.Phys.A,8,(1975),L29. 
\bibitem{4} T.Padmanabhan,The Structure formation in the Universe,(1993)Cambridge University Press.
\bibitem{5} L. C. Garcia de Andrade,Cosmological Density Perturbations in the spin dominated universe and COBE data,submitted to Monthly Notices of the Royal Astronomical Society (2000).
\bibitem{6} B.Kuchowicz,General Relativity and Gravitation Journal (1978),9,6,511.
\bibitem{7} V. de Sabbata and C.Sivaram,Spin and Torsion in Gravitation,(1994),World Scientific.
\end{thebibliography}
\end{document}